\documentstyle[12pt]{article}

\begin{document}

\newcommand{\gi}{g_{i}}
\newcommand{\coor}{X^{ax}}
\newcommand{\coorp}{\dot{X}^{ax}}

\title{Classical Loop Actions of Gauge Theories}

\author{ \\ \\
\small {\bf  D. Armand Ugon, R. Gambini, J. Griego and L. Setaro} \\
\small Instituto de F\'\i sica, Facultad de Ciencias \\
\small                 Trist\'an Narvaja 1674 \\
\small                 Montevideo, Uruguay. }

\date{July 1993}
\maketitle
\vspace{0.5cm}


\begin{abstract}
Since the first attempts to quantize Gauge Theories and Gravity in
the loop representation, the problem of the determination of the
corresponding classical actions has been raised. Here we propose a
general procedure to determine these actions and we explicitly apply
it in the case of electromagnetism. Going to the lattice we show
that the electromagnetic action in terms of loops is equivalent to
the Wilson action, allowing to do Montecarlo calculations in a gauge
invariant way. In the continuum these actions need to be regularized
and they are the natural candidates to describe the theory in a
``confining phase''.
\end{abstract}

\newpage


Loop space provides a common scenario for a  nonlocal  description
of Gauge Theories \cite{Man62},\cite{MaMi81},\cite{Gam86} and
Quantum Gravity \cite{Rov90} ,\cite{Gam91}. Loops form a group, and
all the kinematical properties of gauge theories are imbedded in
this group. This is the remarkable and fundamental property of the
group of loops. Gauge theories arise by considering different
representations  of this group \cite{Ga81},\cite{Ba91}. The Loop
representation \cite{Ga80},\cite{Gam86},\cite{Rov90} is usually
constructed by means of the non canonical algebra of a complete set
of gauge invariant operators. These operators act on  a  state space
of loop wavefunctions $\psi(\gamma)$. Once the complete set of
invariant operators are realized in the space of loops, the action
of any other gauge invariant operator (like the hamiltonian)  can
be  obtained from them. In the abelian case, the noncanonical
algebra is given in terms  of the holonomy $H_A (\gamma)$ and the
conjugate electric field $E^a(x)$.

The non canonical character  of  the  algebra  of  the  fundamental
invariant  operators  shadows   the   classical   counterpart   of
the theory. We do not have at our disposal a  pair  of  canonical
variables whose commutator can be related with a classical Poisson
bracket. As a consequence, we are not able to make  a  Legendre
transform from the hamiltonian to  the  classical  action  of  the
theory. Classical actions in terms of loops would be interesting in
its own  right because they are the natural candidates to describe
the theory in a ``confining phase''. They would also allow to do
Montecarlo calculations in Lattice Gauge Theories in a gauge
invariant way, combining the power of Montecarlo methods with the
advantage of the geometrical character of loop space \cite{Gam86}.
The classical action may be also useful to obtain semiclassical
approximations to the gauge theory under consideration or to General
Relativity in terms of the Ashtekar's variables.

Recently, an extension of the loop space with the structure of  an
infinite dimensional Lie group has been proposed \cite{Gri92}. The
usual group of loops is a subgroup of this extended loop group  and
generalized holonomies can  be  defined  in  the  extended  space.
In this  letter, we  show  that   the   introduction  of   a    new
representation  of  gauge  theories  in  terms  of  this  extended
structure allows to obtain their corresponding  classical  actions
in terms of loops. We shall consider in detail the  electromagnetic
case and sketch the general method for an arbitrary gauge theory. In
a forthcoming paper the extended representation of the Yang Mills
theories and Quantum Gravity will be introduced and the
corresponding classical action will be explicitly calculated. These
representations correspond to the quantized version  of the
classical theories whose actions are given in terms of the elements
of  the extended group. Once the extended classical theory is given,
it is straightforward to specialize them to loops, giving rise to  a
classical  description   of   the theory   in   the confining phase.

The consistency of the loop dependent action for quantum
electromagnetism  will be studied,  showing that  the  loop
dependent classical   action of electromagnetism  in the lattice
leads to the usual Kogut-Susskind hamiltonian.

We start with a brief review of the extended loop group and the
extended loop representation. A more complete treatment  of  these
subjects can be found in references \cite{Gri92}.

The holonomy of a nonabelian connection $A_{ax}$ can be written in
the following way
\begin{equation}
H_A(\gamma)=1+\sum_{n=1}^{\infty} \int dx^3_1\ldots dx^3_n
  A_{a_1}(x_1)\ldots A_{a_n}(x_n)
     X^{a_1\ldots a_n}(x_1,\ldots ,x_n,\gamma) ,
\label{holonomia}
\end{equation}
where
\begin{equation}
X^{a_1\ldots a_n}(x_1,\ldots,x_n,\gamma) = \oint_\gamma
dy_n^{a_n}\ldots \oint_\gamma dy_{1}^{a_1}
   \delta (x_n-y_n)\ldots \delta(x_1-y_1)
               \Theta_\gamma(0,y_1,\ldots,y_n)
\end{equation}
and $\Theta_\gamma(0,y_1,\ldots,y_n)$ orders the points along  the
contour starting at the origin of the loop. These relations define
the  multitangent field  of   rank   $n$  associated  to  the   loop
$\gamma$.  They  behave  as multivector  densities under   general
coordinate transformations.  No  more information  from  the  loop
is  needed  in  order  to  compute  the holonomy than what is
present in the multitangents of all orders.

In order to simplify the notation, it is convenient to use the
following conventions
\begin{equation}
X^{\mu_1 \ldots \mu_n}(\gamma) = X^{a_1\,x_1\ldots a_n\,x_n}(\gamma)
= X^{a_1\ldots a_n}(x_1,\ldots,x_n,\gamma) \;,
\end{equation}
with $\mu_i \equiv (a_ix_i)$. The X's are not independent
quantities, they  obey  two  kinds  of constraints: the algebraic
and differential constraints.

The algebraic constraints stem from the relations satisfied by the
generalized Heaviside function and have the general form
\begin{equation}
\sum_{P_k} X^{P_k(\mu_1 \mu_n)} = X^{\mu_1\ldots\mu_k} \,
                                          X^{\mu_{k+1}\ldots
\mu_n}
\label{VA}
\end{equation}
where the  sum  goes  over  all  the  permutations  of  the  $\mu$
variables which preserve the ordering of  the  ${  \mu_1,  \ldots,
\mu_k  }$  and  the  ${  \mu_{k+1},  \ldots,  \mu_n   }$   between
themselves.

The differential constraints ensure  that  the  holonomy  has  the
correct transformation properties under gauge transformations, and
can be readily derived from Eq. (\ref{holonomia}).  They  are
given by,
\begin{equation}
\frac {\partial \phantom{iii}} {\partial x_i^{a_i}} \,
X^{a_1 x_1\ldots a_i x_i \,\ldots\,a_nx_n} =
 \bigl( \,\delta(x_i-x_{i-1}) - \delta(x_i-x_{i+1}) \,\bigr)
 X^{ a_1 x_1\ldots a_{i-1} x_{i-1}\,a_{i+1} x_{i+1}\ldots a_n x_n}
\label{VD}
\end{equation}
In this expression, points $x_0$ and $x_{n+1}$ are to be understood
as the basepoint of the loop.

An important property  of  the   constraints  is  that  {\em  any}
mutitensor density $X^{a_1 x_1 \ldots  a_n  x_n}$  that  satisfies
them can be used in Eq. (\ref{holonomia}) and the  resulting
object  is  a  {\em    gauge     covariant}     quantity.     When
restricted to the multitangents $X(\gamma)$ associated with loops,
the resulting object is the holonomy. It is  this  property  that
allows to extend the loops to a more general structure.   One  can
in general deal with  arbitrary  multitensor  densities  $X$  (not
necessarily related with  loops)  and  construct  gauge  invariant
objects,  by  taking  the  trace.  The  multitensor
densities  need   not   be   distributional   functions   as   the
multitangents associated with a  loop.  They  could  be  perfectly
smooth functions on the manifold.

With this construction in hand, one could go  further  and  forget
loops and holono\-mies altogether. Since  one  can  represent  any
gauge covariant object using the X's, one  can  represent  a gauge
theory  {\em   entirely}   in   terms   of   X's.  The  underlying
mathematical structure  of  this  extended representation will  be
called the ``extended group of loops'' which  has  the   structure
of an infinite  dimensional  Lie   group.

The extended group can be introduced in the following way: given
a set of arbitrary (unconstrained) multitensor
densities of any rank $E^{\mu_1\ldots \mu_n}$ we define the object
${\bf E}$ as
\begin{equation}
{\bf E} \; = \; (\, E, \, E^{\mu_1}, \, \ldots , \,
E^{\mu_1\ldots \mu_n}, \, \ldots ) \; \equiv (\, E, \, \vec{E})
\end{equation}
where $E$ is a real number. It can  be  readily checked  that  the
set $\{ {\bf E} \}$ has the structure of a vector  space  (denoted
as ${\cal E}$) with the  usual  composition   laws   of   addition
and multiplication.

We will now introduce a product law in ${\cal E}$ in the following
way: given two vectors ${\bf E}_1$ and ${\bf E}_2$, we define ${\bf
E}_1 \times {\bf E}_2$ as the vector with components
\begin{equation}
{\bf E}_1 \times {\bf E}_2 \; = \; (\,E_1 E_2, \, E_1 \vec{E_2}
\, + \, \vec{E_1}
E_2 \, + \, \vec{E_1} \times \vec{E_2})
\end{equation}
where  $\vec{E_1}  \times  \vec{E_2}$   is   given   by,
\begin{equation}
(\vec{E}_1  \times  \vec{E}_2)^{\mu_1\ldots \mu_n}
=   \sum^{n-1}_{i=1}
       E_1^{\mu_1\ldots \mu_i}\,E_2^{\mu_{i+1}\ldots \mu_n}\;\;.
\end{equation}

The product law is associative and distributive  with  respect  to
the addition of vectors. It has a null element (the  null  vector)
and a identity element, given by
\begin{equation}
{\bf I} = (\, 1,\, 0,\, \ldots ,\, 0, \, \ldots) \;\; .
\end{equation}
An inverse element exists for all vectors with nonvanishing zero
rank component. It is given by
\begin{equation}
{\bf E}^{-1} \; = \; E^{-1} {\bf I} \; + \; \sum^{\infty}_{i=1}
(-1)^i E^{-i-1}
({\bf E} \, - \, E {\bf I})^i \label{inverse}
\end{equation}
such that
\begin{equation}
{\bf E} \times {\bf E}^{-1} \; = \; {\bf E}^{-1} \times {\bf E}
 \; = \; {\bf I} \;\; .
\end{equation}

The set of all vectors with nonvanishing zero rank component (notice
the $E^{-1}$ role in Eq. (\ref{inverse})) forms a group with the
$\times$-product law.

Introducing supplementary conditions related to the  constraints, it
is possible to define several  base pointed  subgroups  of  this
general group. Among these  we  find  the Extended Loop group,
defined as the set ${\cal D}_{o}$ of elements ${\bf X}=(X,{\vec X})$
with $X$ a nonzero real number and where the  multivector components
$X^{\mu_1  \ldots  \mu_n}$ satisfy the  differential constraint
(\ref{VD}) for any rank $n$.

The group of loops ${\cal L}_{o}$ is a subgroup of the  group
${\cal D}_{o}$
since any multitangent field ${\bf X}(\gamma) \in {\cal D}_{o}$
and
\begin{equation}
{\bf X}(\gamma_{1} \circ \gamma_{2})= {\bf X}(\gamma_{1})\times
{\bf X}(\gamma_{2})
\end{equation}
 therefore ${\cal L}_{o} \subset {\cal D}_{o}$.

Matrix representations of these groups may be  generated   through
a natural extension  of   the   holonomy   (\ref{holonomia}).  The
extended holonomies  associated  with   a    general    connection
$A_{ax}$ are defined by
\begin{equation}
H_A({\bf X})= \sum^{\infty}_{n=0}
 \, i^n \, A_{\mu_1 \ldots \mu_n}
X^{ \mu_1 \ldots \mu_n}
\end{equation}
where a   generalized   Einstein   convention   was   assumed  and
$X^{\mu_1 \ldots \mu_0} = X$. It is straightforward to see that
the generalized holonomies satisfy
\begin{equation}
H_A({\bf X}_1) \, H_A({\bf X}_2) \,=\, H_A({\bf X}_1 \times {\bf
X}_2)
\end{equation}

Let us now recall how the loop representation is usually derived in
the simplest case of the electromagnetic theory.

One  starts  by  considering   the   non
canonical algebra of a complete set of gauge  invariant  operators.
In the Maxwell case \cite{No82}, \cite{Ash92} this  algebra  is
defined in  terms  of  the
gauge invariant holonomy
\begin{equation}
{\hat H}(\gamma)=\exp {i\oint_{\gamma}{\hat A}_{a}(y)dy^{a}}
=\exp {i\int d^3x A_a(x) X^{ax}(\gamma)}
\end{equation}
where
\begin{equation}
X^{ax}(\gamma)= \oint_{\gamma} dy^a
\delta(x-y)
\end{equation}
and the conjugate electric field  ${\hat E}^{a}(x)$. These
operators satisfy the following commutation
relation
\begin{equation}
[{\hat E}^{a}(x),{\hat H}(\gamma)]
\equiv X^{ax}(\gamma) {\hat H}(\gamma)
\label{loopal}
\end{equation}
In the abelian case the theory is completely described in terms of
the the rank one component of the multitangent fields \footnote{This
is true if one considers the restriction of the theory to the
subgroup ${\cal X}_o \subset {\cal D}_o$, where the elements of
${\cal X}_o$ satisfy the differential {\it and} the algebraic
constraints.}.
The gauge invariant operators act  on  a  state  space
of  abelian loop functions  $\psi(\gamma)$  that  may  be  related
with the states in the connection  representation by the loop
transform
\begin{equation}
\psi(\gamma)=\int d_{\mu}[A] \, \psi[A] \, H_A ({\gamma})
\end{equation}
By means of the loop transform it
is immediate to calculate the  explicit action  of  the fundamental
gauge invariant operators
\begin{eqnarray}
&&{\hat H}(\gamma_{0})\psi(\gamma)=\psi(\gamma_{0}\circ\gamma) \\
&&{\hat E}^{a}(x)\psi(\gamma) =
      \oint_{\gamma}\delta(x-y)dy^a \psi(\gamma)
\end{eqnarray}
The  action  of  any  other  gauge  invariant
operator   in   the abelian loop representation may be deduced from
the above expressions. For  instance the hamiltonian takes the form
\begin{equation}
{\hat H}\psi(\gamma)=[\int{d^{3}x\Delta_{ij}(x)\Delta_{ij}(x)}+
\oint_{\gamma}dy^{a}\oint_{\gamma}dy'^{a}\delta^{3}(y-y')]
\psi(\gamma)
\end{equation}
where $\Delta_{ij}(x)$ is the loop
derivative \cite{Gam86}. The  electric  term of the hamiltonian is
singular  and  needs  to  be  regularized. One usually introduces a
regularization  of  the  $\delta$-function, for instance
\begin{equation}
f_{\epsilon}(x-y)= (\pi \epsilon)^{-3/2}\exp(-(x-y)^{2}/  \epsilon)
\end{equation}
Similar
regularizations   are   required   for   the   loop representations
of  the  nonabelian  gauge  theories  and  quantum gravity, and
they are also  required  for  other  descriptions  in terms of
loops in the continuum. We shall see that the extended loop
representation  is  free  from singularities of this kind.

We now turn to the extended representation. The extended
representation of an abelian gauge theory is based in the following
extension of the holonomy
\begin{equation}
H_A (\gamma) \rightarrow  H_A ({\bf X}) =
\exp {ig \int d^3 x A_{ax}\,X^{ax}}
\end{equation}
where we have explicitly introduced the coupling constant $g$ and
$X^{ax}$ is an arbitrary divergence free vector density field
\begin{equation}
\partial_{ax}X^{ax} = 0
\end{equation}
that correspond to the first component of the multivector densities
${\vec X}$. Notice that now the fields $X^{ax}$ are not necessarily
related with a loop, they may be smooth functions of the space
coordinates.
The new non canonical gauge invariant algebra
\begin{equation}
[{\hat E}^{ax}, {\hat H}_{A}({\bf X})]
           = g\,X^{ax} {\hat H}_{A}({\bf X})
\end{equation}
can be realized on the linear space of extended abelian loop
functions $\psi(X^{ax})$. The extended wavefunctions will be
connected with the states in the connection representation by means
of a generalized loop transform
\begin{equation}
\psi({\bf  X})=\int  d_{\mu}[A]  \psi[A]   \exp{(-ig\int   d^3x
A_a(x)X^{ax})} \label{transf2}
\label{transext}
\end{equation}

The operators ${\hat H}_{A}({\bf X})$ and ${\hat E}^{ax}$
are realized in the extended space, as was done before in the loop
representation. We get
\begin{eqnarray}
&&{\hat H}_{A} ({\bf X}_{0}) \, \psi ({\bf X}) = \psi ({\bf X} +
{\bf  X}_{0})\\
&&{\hat E}^{ax} \, \psi ({\bf X}) = g X^{ax} \psi ({\bf X})
\label{electrico}
\end{eqnarray}
The action of any other gauge invariant operator may be again
deduced from them. The magnetic field operator takes the form
\begin{equation}
\label{mag}
{\hat F}_{ab}(x) \, \psi ({\bf X}) = \frac{i}{g} \,
\partial_{[a} \frac{\delta}
{\delta X^{b]x}} \, \psi ({\bf X})
\end{equation}
and the hamiltonian operator is given by
\begin{equation}
{\hat{\cal  H}}   \,   \psi   ({\bf    X})    =     \int    d^{3}x
  \left[ \,
\frac{g^2}{2}{\hat X}^{ax} {\hat X}^{ax} + \, \frac{1}{4g^2} \, (
\partial_{[a} {\hat P}_{b]x} )^2 \right] \, \psi({\bf X})
\label{hham}
\end{equation}
where
\begin{equation}
{\hat P}_{bx} = i \, \frac{\delta}{\delta X^{bx}}
\end{equation}

In the abelian case the extended representation is nothing but an
electric  field representation of electromagnetism. It is
straightforward to see, that using (\ref{electrico}) and (\ref{mag})
and observing that $g\hat{A}_a(x) = \hat{P}_a(x) $, one reobtains
the standard form of the hamiltonian.

The transforms (\ref{transf2}) and (\ref{transext}) allow to define
a one to one correspondence between loop dependent solutions
$\psi(\gamma) = \psi[X(\gamma)] $ and the extended solutions
$\psi({\bf X})$. The correspondence arises simply by substituting
$X(\gamma) $ by ${\bf X}$. For instance the vacuum of the
electromagnetic theory in the loop representation takes the form

\begin{equation}
\psi(\gamma) = \exp{ -({\textstyle{g^2 \over2}}) \int_{\gamma} dy^a
\int_{\gamma} dy^{a'} D_{1}(y-y') }
\end{equation}
where $D_1$ is the homogeneous symmetric propagator, while the
corresponding state in the extended representation is
\begin{equation}
\psi({\bf X}) = \exp{ -({\textstyle{g^2 \over2}})
                            \int d^3 x \int d^3 y X^{ax}
X^{ay} D_{1}(x-y)}
\end{equation}
where now $X^{ax}$ is a general divergence free object. While the
exponent of the loop dependent state is singular and needs to be
regularized the wave function in the extended space is perfectly
well defined when evaluated on smooth functions. An analogous
situation appears in the non abelian gauge theories and quantum
gravity where the use of the extended loop representation allows to
get rid of ambiguities introduced by the regularization procedure.
As shown by Ashtekar and Isham \cite{Ish91}, the loop algebra
(\ref{loopal}) has several kinds of representations. In  loop space,
it is possible to introduce an inner product in terms of the $X$
variables that allows to get a Fock representation of the theory.
This representation gives a correct description of photons or
gravitons in the linearized gravity case,  which amounts to consider
implicitly the extended representation. One can also consider
another inequivalent representation which is defined in the space of
functions of loops which are normalizable with respect to a discrete
measure. This new Fock representation seems to describe the quantum
states of a Type II superconductor  and it should be considered as
the most natural representation when one works with true loops.

Notice that once we have the quantum form of the hamiltonian written
in terms of the elements of the extended group (in this case in
terms of the elements of the abelian extended group) it is
straightforward to write a classical action associated  with the
theory. The classical action associated with the hamiltonian
(\ref{hham}) is
\begin{equation}
S = \int dt \left\{ \, P_{ax} \dot{X}^{ax} -
    \left[ \frac{g^2}{2} \, X^{ax} X^{ax} +
      \frac{1}{4g^{2}} \, ( \partial_{[a}P_{b]x} )^2 \right]
                                + \lambda_x X^{ax}_{,a} \right\}
                                                   \label{action}
\end{equation}
that may be simply recognized as the usual action of
electromagnetism if we impose
\begin{equation}
gX^{ax} = E^a (x) \;\; ; \;\;\;\; \partial_{ax} E^a (x) = 0
\end{equation}
and
\begin{equation}
P_{ax} = g A_a (x) \;\; ; \;\;\;\; \lambda_x = A_0 (x)
\end{equation}
When this action is restricted to loops
\begin{equation}
X^{ax} = X^{ax} (\gamma) = \int_{\gamma} dy^a \delta (x - y)
\end{equation}
one gets
\begin{equation}
S =  \int dt \left\{ \, \int_{\gamma_t} dy^a \dot{A}_{a} (y) +
{\textstyle{1\over2}} \, \int d^3 x F_{ab} (x) F_{ab} (x) +
g^2\int_{\gamma_t} dy^a \int_{\gamma_t} dy'^a f_{\epsilon} (y-y') \,
\right\}
\end{equation}
where $f_{\epsilon}$ is a regularization of the $\delta$  function
and the loops $\gamma_t$ belongs to the surface $t=$ constant.

We may go to the second order form of the action. We get, ignoring
the regularization
\begin{equation}
\label{lagrange}
S = {\textstyle{g^2\over2}} \int dt \int d^3{x} \int d^3{y}
\left\{ \, -\dot{X}^{ax} (\gamma)
            \frac{1}{4\pi \mid x-y \mid} \dot{X}^{ay} (\gamma)
 - X^{ax} (\gamma) \, \delta(x-y) \, X^{ay} (\gamma) \, \right\}
                                                   \label{action2}
\end{equation}
where
\begin{equation}
\dot{X}^{ax} (\gamma) = \lim_{\delta t \rightarrow 0}
      \frac{X^{ax} (\gamma_{t+\delta t})
              -  X^{ax}  (\gamma_{t})}{\delta t}
\end{equation}

As we have already noticed, this action is singular in the continuum
and needs to be regularized. In the lattice we shall see that it is
perfectly well defined and leads, via   the transition matrix, to
the usual loop representation of electromagnetism.

The extended representation for a nonabelian gauge theory may be
obtained following a similar procedure as in the abelian case. In
the nonabelian case the wavefunctions $\psi({\bf X})$ defined on the
extended group that satisfies the differential constraint ${\cal
D}_o$ will depend on the multivector component ${\vec X}$ of any
rank. The hamiltonian may be written in terms of $X^{\mu_1 \ldots
\mu_n}$ and its conjugate momentum $P_{\mu_1 \ldots \mu_n}$, given
by
\begin{equation}
P_{\mu_1 \ldots \mu_n} = i \,
          \frac{\delta}{\delta X^{\mu_1 \ldots \mu_n}}
\end{equation}
{}From the hamiltonian on can read the corresponding classical
action written in terms of the canonical variables $(X^{\mu_1 \ldots
\mu_n},P_{\mu_1 \ldots \mu_n})$.  The restriction of this extended
action to multitangent fields
\begin{equation}
X^{\mu_1 \ldots \mu_n} = X^{\mu_1 \ldots \mu_n}(\gamma)
\end{equation}
leads to an action in terms of loops for the Yang Mills theory. The
physical meaning of this action is still unclear, but is natural to
expect that it might be relevant  to the semiclassical description
of the physical excitations in the confining phase.

To prove the consistency of this  approach we will now quantize the
classical action (\ref{action2}) in the lattice using the transfer
matrix formalism \cite{Kog79}, \cite{Cre}. We  will  first  proceed
to   obtain a   lattice   version   of (\ref{lagrange}). In a cubic
lattice with $N^3$ sites the coordinates become \begin{equation}
x_{i} = a \, n_{i} \end{equation} where $a$ is the lattice  spacing
and $- \frac{N}{2} < n_{i} < \frac{N}{2}$

The Laplacian operator can be easily inverted using the Fourier
transform. The discrete Fourier transform of $\dot{X}^{ax}$ is
\begin{equation}
\dot{X}^{a} (\vec{n}) =
  \frac{1}{N^{3} a^{3}}
    \sum_{\vec{q}} e^{-i \vec{q} \cdot \vec{n} a}
       \stackrel{\sim}{\dot{X}^{a}} (\vec{q})
\end{equation}
and its inverse
\begin{equation}
\stackrel{\sim}{\dot{X}^{a}} (\vec{q}) =
   \sum_{\vec{n}} e^{i \vec{q} \cdot \vec{n} a}
                                  \dot{X}^{a} (\vec{n})
\label{if}
\end{equation}
where the $\vec{q}$  components run over the first Brioullin zone
$- \frac{\pi}{a} < q_{i} \leq \frac{\pi}{a}$. Using the discrete
form of the Laplacian operator we get
\begin{equation}
\frac{1}{\nabla^2}
             \dot{X}^{a} (\vec{n}) =
     - \sum_{\vec{n}'} f(\vec{n} - \vec{n}')
                         \dot{X}^{a} (\vec{n}')
                                          \label{Xpunto}
\end{equation}
where
\begin{equation}
f(\vec{n} - \vec{n} ') = \frac{a^{2}}{2 N^{3}}
  \sum_{\vec{q}}
   \frac{e^{-i \vec{q} \cdot (\vec{n}-\vec{n} ') a}}
       {\sum_{i=1}^{3} \left[1 - cos(q_{i} a) \right]}
\end{equation}

The temporal derivative of $X^{ax}$ is given by
\begin{equation}
\dot{X}^{a}_{\gamma} (\vec{n}) = \frac{1}{\tau}
     \left( X^{a}_{\gamma'} (\vec{n}) -
                         X^{a}_{\gamma} (\vec{n}) \right)
                = \frac{1}{\tau}X^{a}_{\gamma'-\gamma}(\vec{n})
\end{equation}
where $\gamma$ and $\gamma'$ are loops in the spatial surfaces  at
the times $t$ and $t + \tau$, and \mbox{$\gamma'-\gamma  =
\gamma'\circ \overline{\gamma}$} where $\overline{\gamma}$ is the
loop  $\gamma$ with  the  opposite  orientation. Notice that this
result  holds because the ``loop coordinates'' $X^{ax} (\gamma)$ are
linear functionals of  loops in the abelian case.

On the lattice, the first term of the classical electromagnetic
action reads
\begin{equation}
 \int d^{3}x \, \dot{X}_{\gamma}^{ax}
    \frac{1}{\nabla^{2}} \dot{X}_{\gamma}^{ax}
       \longrightarrow
\frac{a^{3}}{\tau^{2}}
         \sum_{\vec{n}} \sum_{\vec{n}'}
         f(\vec{n} - \vec{n}') X^{a}_{\gamma'-\gamma} (\vec{n})
                                X^{a}_{\gamma'-\gamma} (\vec{n}')
\end{equation}
with
\begin{equation}
X^{v}_{\gamma}(\vec{n}) = \frac{1}{a^2} \sum_{\ell ' \in \gamma}
                          \overline{\delta}_{\ell \ell '}
\end{equation}
where $\ell$ is the link with origin $\vec{n}$ and orientation $v$,
and
\begin{equation}
\overline{\delta}_{\ell {\ell '}}
               = \left\{ \begin{array}{rl}
                      1   & \mbox{for $\ell = \ell '$} \\
                     -1   & \mbox{for $\ell = \overline{\ell '}$} \\
                      0   & \mbox{otherwise}
                        \end{array}
                 \right.
\end{equation}
The second term takes the form
\begin{equation}
\int d^{3}x \, X_{\gamma}^{ax} X_{\gamma}^{ax} \longrightarrow
   \frac{1}{a} \sum_{\ell \in \gamma} \sum_{\ell ' \in \gamma}
       \overline{\delta}_{\ell    \ell    '}     =     \frac{1}{a}
\Lambda_{\gamma}
                                                  \label{Lagr-2}
\end{equation}
where $\Lambda_{\gamma}$ is the quadratic length of the
loop $\gamma$.
Doing a Wick rotation, we obtain for the  euclidean lattice action
the following expression
\begin{equation}
S_{E} = K_{\tau}  \sum_{n_{0}} D(\gamma'-\gamma)
        + K \sum_{n_{0}} \Lambda_{\gamma}
                                                  \label{Se}
\end{equation}
where $n_{0}$ is the discrete time variable,
$\gamma$ and $\gamma'$ are loops in the spatial
surfaces at the times $n_{0} \tau$ and $(n_{0} + 1) \tau$, and

\begin{equation}
K_{\tau} = \frac{g^{2}_{\tau} a^{3}}{2  \tau} \;\;\; , \;\;\;
                     K = \frac{g^2 \tau}{2a}
\end{equation}
where $g_{\tau} = g$ when $\tau = a$ and with
\begin{equation}
D(\gamma'-\gamma) = \sum_{\vec{n}} \sum_{\vec{n}'}
        f(\vec{n} - \vec{n}')
         X^{a}_{\gamma'-\gamma}(\vec{n})
                 X^{a}_{\gamma'-\gamma}(\vec{n}')
                                                \label{D}
\end{equation}

Let us now introduce the partition function
\begin{equation}
Z = \int_{- \infty}^{+ \infty}
         \left[ dX \right] e^{-S_{E}}
\end{equation}
where the integral is done  over  all   the   configurations   of
the abelian loop coordinates $X^{ax}(\gamma)$. Since  the
euclidean   action in  the  lattice  only involves the  loop
coordinates  at  $n_{0} \tau$  and  $(n_{0}+1)\tau$,  we  can  use
the  transfer   matrix formalism and write  the  partition function
as
\begin{equation}
Z =  \int_{- \infty}^{+ \infty} \left[ dX \right]
     \prod_{n_{0}}  \langle  X_{\gamma'}  \mid  \hat{T}   \mid
X_{\gamma} \rangle
\end{equation}
where as before $\gamma$ and $\gamma'$ are loops in the spatial
surfaces at  the  times  $n_{0}  \tau$  and  $(n_{0}  +  1)  \tau$
respectively.

The operator $\hat{T}$
acts in the space of loop coordinates and its matrix elements are
\begin{eqnarray}
\label{transfer1}
 \langle X_{\gamma'} \mid &\hat{T}& \mid X_{\gamma} \rangle
 =  T(\gamma', \gamma) \nonumber \\
&=&
  exp \left( -K_{\tau}  \sum_{\vec{n}}
         \sum_{\vec{n}'} f( \vec{n} - \vec{n'})
           X^{a}_{\gamma'-\gamma}(\vec{n})
                  X^{a}_{\gamma'-\gamma}(\vec{n'})
        - K \Lambda_{\gamma} \right)
 \end{eqnarray}
For a small temporal lattice spacing $\tau$ we have
\begin{equation}
\hat{T} = e^{- \tau \hat{H}} \approx 1 - \tau \hat{H}
                                            \label{T-H}
\end{equation}
where $a$ is held fixed and $\hat{H}$ is the hamiltonian
operator formulated on a spatial lattice.

Next we must determine how $K_{\tau}$ and $K$ should be adjusted so
that the transfer matrix elements take the form (\ref{transfer1}).
The diagonal elements  of  the  matrix  $  T(\gamma',\gamma)$  are
obtained when the loops at times $n_{0} \tau$ and $(n_{0}+1) \tau$
are the same (that is to say, $\gamma'-\gamma$  is  equivalent  to
the null path)
\begin{equation}
T (\gamma,\gamma) =
  e^{-K   \Lambda_{\gamma}}   \approx   1   -   \tau   \hat{H}
\mid_{\gamma,\gamma}
                                               \label{T00}
\end{equation}
As we shall see below, the largest contribution of the off
diagonal terms is obtained when the difference between the loops at
times $n_{0} \tau$ and $(n_{0}+1) \tau$ is a plaquette with positive
or   negative   orientation   ($\gamma'-\gamma    =    \Box$    or
$\gamma'-\gamma = \stackrel{-}{\Box}$)
\begin{equation}
T (\gamma \circ \Box,\gamma)  =
  e^{-K_{\tau} D(\Box)} e^{-K \Lambda_{\gamma}}
      \approx -\tau \hat{H} \mid_{\gamma \circ \Box,\gamma}
                                                \label{T10}
\end{equation}
\begin{equation}
\hat{T} (\gamma \circ \stackrel{-}{\Box},\gamma)  =
  e^{-K_{\tau} D(\Box)} e^{-K \Lambda_{\gamma}}
    \approx     -\tau     \hat{H}      \mid_{\gamma      \circ
\stackrel{-}{\Box},\gamma}
                                                \label{T10b}
\end{equation}
For the case of $\gamma'-\gamma    =    \Box$ the sum (\ref{D}) is
reduced to
\begin{equation}
D(\Box) = \frac{2  (N+1)^{3}}{3 a^{2} N^{3}}
\end{equation}
and for $N \gg 1$ one can see (for the (2+1)-dimensional case) that
\begin{equation}
D(\gamma'-\gamma) \approx m \ D(\Box)
\label{aprox}
\end{equation}
where $m$ is  the  number  of  plaquettes  of  the  loop
$\gamma'-\gamma$. In the (3+1)-dimensional case we get
\begin{equation}
\label{area}
D(\gamma'-\gamma) \approx L
\label{aproxL}
\end{equation}
where $L$ is the length of the loop $\gamma ' - \gamma$. From Eqs.
(\ref{T00}), (\ref{T10}) and (\ref{T10b}) we see that the following
conditions must be required

\begin{eqnarray}
\label{tau1}
K                      & \approx & \tau \nonumber \\
e^{- K_{\tau} D(\Box)} & \approx & \tau
\end{eqnarray}
The others matrix elements do  not  survive  since  according  to
(\ref{aprox}) or (\ref{area})
\begin{equation}
T (\gamma',\gamma)  =
  e^{-K_{\tau} D(\gamma'-\gamma)} e^{-K \Lambda_{\gamma}}
  \approx  e^{-m K_{\tau} D(\Box)} e^{-K
\Lambda_{\gamma}} = O(\tau^{m})
                                                \label{TLL}
\end{equation}
Then  from Eq. (\ref{tau1})  we must have
\begin{eqnarray}
K = \lambda \, e^{- K_{\tau} D(\Box)}
\end{eqnarray}
If we identify the temporal lattice spacing as
\begin{eqnarray}
\tau = a \, e^{- K_{\tau} D(\Box)}
\end{eqnarray}
we find
\begin{equation}
K = \frac{\lambda}{a} \, \tau
\end{equation}
{}From Eqs. (\ref{T00}), (\ref{T10}) and (\ref{T10b}) we conclude
that
\begin{eqnarray}
\langle \gamma \mid \hat{H} \mid \gamma \rangle & = &
                         \frac{\lambda}{a} \, \Lambda_{\gamma}
                         \nonumber \\
\langle \gamma \circ \Box \mid \hat{H} \mid \gamma \rangle
                                      & = & - \frac{1}{a} \\
\langle \gamma \circ \stackrel{-}{\Box}
               \mid \hat{H} \mid \gamma \rangle & = & - \frac{1}{a}
               \nonumber
\end{eqnarray}
{}From the above results, the following expression for
the hamiltonian operator in the lattice is obtained
\begin{equation}
{\hat H} = - \frac{1}{a} \sum_{\Box} \left( W(\Box) +
            W^{\dagger}(\Box) \right)
+ \frac{\lambda}{a} \, {\hat E}
                                     \label{H}
\end{equation}
where the action of $W(\gamma) $ and ${\hat E}$ are defined by the
following expressions
\begin{eqnarray}
W(\gamma') \mid \gamma \rangle     & = &
                \mid  \gamma' \circ \gamma \rangle \nonumber \\
W^{\dagger}(\gamma') \mid \gamma \rangle & = &
     \mid \overline{\gamma'} \circ \gamma \rangle \\
{\hat E} \mid \gamma \rangle    &=&
                    \Lambda_{\gamma} \mid \gamma \rangle
                    \nonumber
\end{eqnarray}

The  hamiltonian (\ref{H}) is  nothing  else  that  the familiar
Kogut-Susskind  hamiltonian  for electromagnetism  in   the   loop
representation. therefore we have shown that the electromagnetic
loop action when quantized in the lattice leads to the usual
Kogut-Susskind.

We conclude with some general remarks. The principal objective of
this letter was to show how a classical action in terms of loops can
be derived for electromagnetism and implemented in the lattice. This
loop action when quantize in the lattice leads to the usual
Kogut-Suskind hamiltonian. We have now at our disposal  a gauge
invariant action, that may serve as the starting point of Montecarlo
calculations of electromagnetism in terms of loops. One has  the
combined  power of montecarlo methods with  the geometric advantages
of the loop representation. In particular, all the techniques
developed in the hamiltonian loop representation of gauge theories
\cite{Lattice} can now be applied in the statistical approach. In
the case of 2+1 dimensions, the loop action has only two terms, one
proportional to the length of the loop and the other to the
difference area of the loops at time $t$ and $t+\tau$. Moreover,
when one considers the regularized version of this action in the
continuum, it does not lead to the usual Maxwell equations but to
the equations for the electromagnetic interactions among confined
electric or magnetic lines of flux in a Type II superconductor. We
wish to thank, A. Aroca, C. Di Bartolo, H. Fort and J. Pullin for
useful discussions and comments.


\end{document}